\documentclass[twocolumn]{apj}

\usepackage{graphicx}	
\usepackage{amsmath}	
\usepackage{amssymb}	
\usepackage{newtxtext,newtxmath}
\usepackage[T1]{fontenc}
\usepackage{ae,aecompl}
\usepackage{color}
\usepackage{ulem}
\usepackage{xspace}
\usepackage{graphicx}
\usepackage{multirow}

\received{January 1, 2018}
\revised{January 7, 2018}
\accepted{\today}
\submitjournal{ApJ}

\shorttitle{Neutrino Losses in Type I X-Ray Bursts}
\shortauthors{Goodwin et al.}

\newcommand{\Qnuc}{{\ensuremath{Q_\mathrm{nuc}}}\xspace}
\newcommand{\Xb}{{\ensuremath{\overbar{X}}}\xspace}
\newcommand{\E}[1]{{\ensuremath{\times10^{#1}}}}

\newcommand{\overbar}[1]{\mkern 1.5mu\overline{\mkern-1.5mu#1\mkern-1.5mu}\mkern 1.5mu}

\newcommand{\KEPLER}{\textsc{Kepler}\xspace}

\newcommand{\MeVn}{{\ensuremath{\mathrm{MeV}\,\mathrm{nucleon}^{-1}}}\xspace}
\newcommand{\MeV}{{\ensuremath{\mathrm{MeV}}}\xspace}
\newcommand{\keVn}{{\ensuremath{\mathrm{keV}\,\mathrm{nucleon}^{-1}}}\xspace}

\begin{document}

\title{Neutrino Losses in Type I Thermonuclear X-Ray Bursts:
An Improved Nuclear Energy Generation Approximation}

\correspondingauthor{Adelle Goodwin}
\email{adelle.goodwin@monash.edu}

\author{A.~J.\ Goodwin}
\affil{School of Physics and Astronomy, Monash University, Clayton, Vic 3800, Australia}
\affil{Monash Centre for Astrophysics}

\author[0000-0002-3684-1325]{A.\ Heger}
\affil{School of Physics and Astronomy, Monash University, Clayton, Vic 3800, Australia}
\affil{Monash Centre for Astrophysics}
\affiliation{Tsung-Dao Lee Institute, Shanghai 200240, China}

\author{D.~K.\ Galloway}
\affil{School of Physics and Astronomy, Monash University, Clayton, Vic 3800, Australia}
\affil{Monash Centre for Astrophysics}
\nocollaboration

\begin{abstract}

Type I X-ray bursts are thermonuclear explosions on the surface of accreting neutron stars.  Hydrogen rich X-ray bursts burn protons far from the line of stability and can release energy in the form of neutrinos from $\beta$-decays.  We have estimated, for the first time, the neutrino fluxes of Type I bursts for a range of initial conditions based on the predictions of a 1D implicit hydrodynamics code, \KEPLER, which calculates the complete nuclear reaction network.  We find that neutrino losses are between $6.7\E{-5}$ and $0.14$ of the total energy per nucleon, \Qnuc, depending upon the hydrogen fraction in the fuel. These values are significantly below the $35\,\%$ value for neutrino losses often adopted in recent literature for the \textsl{rp}-process. The discrepancy arises because it is only at $\beta$-decays that $\approx35\,\%$ of energy is lost due to neutrino emission, whereas there are no neutrino losses in $(p,\gamma)$ and $(\alpha,p)$ reactions.  Using the total measured burst energies from \KEPLER for a range of initial conditions, we have determined an approximation formula for the total energy per nucleon released during an X-ray burst, $\Qnuc=(1.31+6.95\,\Xb-1.92\,\Xb^2)\,\MeVn$, where $\Xb$ is the average hydrogen mass fraction of the ignition column, with an RMS error of $0.052\,\MeVn$. We provide a detailed analysis of the nuclear energy output of a burst and find an incomplete extraction of mass excess in the burst fuel, with $14\,\%$ of the mass excess in the fuel not being extracted.  

\end{abstract}

\keywords{}



\section{Introduction}\label{sec:intro}

Since the discovery of bright, energetic explosions from compact objects, known as thermonuclear X-ray bursts \citep{Grindlay1976}, observers have used these events as laboratories for nuclear physics experiments that cannot be replicated on Earth.  Type I X-ray bursts are highly energetic ($\sim10^{38}\,$erg) thermonuclear flashes observed radiating from the surface of an accreting neutron star \citep[e.g.,][]{lewin1993}.  They are the most frequently observed thermonuclear-powered outbursts in nature and can be used to constrain fundamental information about matter of super-nuclear density and the nuclear reactions and processes that can occur in extreme environments.

Thermonuclear X-ray bursts occur at high ($>10^7\,$K) temperatures.  They are powered by nuclear reactions, with current theory suggesting five main nuclear reaction pathways involved in the burning in the lead up to a burst, as well as during a burst.  These are: the (hot) CNO cycle, the $3\alpha$-process, the $\alpha$-process, the $\alpha$\textsl{p}-process, and the \textsl{rp}-process \citep[e.g.,][]{galloway2017,bildsten1998}.

The hot CNO cycle burns hydrogen into helium.  The process involves two $\beta$-decays, releasing 2 neutrinos which carry away a total of $\sim2\,$MeV, and is usually responsible for the steady hydrogen burning between bursts, and could be responsible for burst ignition at very low accretion rates \citep{fujimoto1981}. 
The triple-$\alpha$ process is the pathway by which helium burns to carbon and is thought to be the primary cause of burst ignition \citep[e.g.,][]{Joss1978}.
The $\alpha$-process forms heavier elements from the products of the previous two processes and occurs at temperatures $\gtrsim10^9\,$K \citep{fujimoto1981}.  The $\alpha$\textsl{p}-process is similar to the $\alpha$-process, but the reaction is catalysed by protons.
The \textsl{rp}-process (rapid-proton capture process) is a chain of successive proton captures and $\beta$-decays, beginning with the products of the previous nuclear reaction chains, and quite easily producing elements beyond $^{56}$Fe \citep[see][]{wallace1981}.  Each $\beta$-decay releases a neutrino.

The primary source of neutrinos during a burst is the $\beta$-decays, where typically $\approx35$\,\% of energy is lost as neutrinos \citep{fujimoto1987}, depending on weak strength distribution, and can be quite different for electron captures.  In recent X-ray burst literature, this is often misinterpreted to mean neutrino energy release for the entire \textsl{rp}-process is $35\,\%$ \citep[e.g.,][]{zand2017,cumming2000}.  To emphasise this point, again, it is only for the $\beta$-decays that $35\,\%$ of the energy is lost as neutrinos and $\beta$-decays do not dominate the total energy release. 

Modelling of Type I thermonuclear X-ray bursts has played a significant role in our understanding of their observational properties.  Understanding of Type I X-ray bursts must combine theory and observations to comprehensively predict observed burst parameters.  Early models \citep[e.g.,][]{fujimoto1981,taam1980,vanparadijs1988,cumming2000}, focus on the correlation between burst energies and recurrence times, to varying success, using an analytic or semi-analytic approach to integrate an ignition column using simple assumptions.  More recent models \citep[e.g.,][]{woosley2004,Fisker2008,Fisker2004} have a heavier focus on the nuclear physics driving the bursts, predicting fuel compositions and accretion rates by implementing a deeper understanding of the nuclear reactions that produce the observed energy generation rates.  The most advanced of these models, \KEPLER \citep{woosley2004}, uses an adaptive nuclear reaction network to model the burning before, and during a burst in more detail. 

Based on an incorrect expression for neutrino losses during the $\alpha$\textsl{p}- and \textsl{rp}-process, the energy generation of Type I bursts (\Qnuc) in simple models has often been given by the relation $\Qnuc=(1.6+4\,\Xb)\,\MeVn$, where \Xb is the average hydrogen mass fraction of the ignition column \citep[e.g.,][]{cumming2000}. The formula accounts for helium burning to iron group with an energy release of $1.6\,\MeVn$ and hydrogen burning to iron group with an energy release of $4\,\Xb\,\MeVn$.  \Qnuc is directly related to the energy of a burst by multiplying by the number of nucleons (mass) burnt in the burst.  As such, this expression for \Qnuc can be used to infer the energy of a simulated burst, given some composition and accretion rate (and hence ignition depth) in simple analytic models \citep[such as][]{cumming2003}.  Observers also use this approximation to infer the average hydrogen fraction at ignition, and thus composition of the fuel burnt in the burst from observations of energy \citep[e.g.,][]{galloway2006,Chenevez2016,galloway2008,galloway2004b}.

In this work we measure the neutrino fluxes of Type I X-ray bursts based on the predictions of the advanced modelling code \KEPLER \citep{woosley2004}, and develop a new nuclear energy generation approximation using these neutrino estimates.  We also examine the metallicity dependence on the neutrino losses and the incomplete burning of the burst fuel.  In Section~\ref{sec:methods} we outline the methods and describe the \KEPLER code used in this work, in Section \ref{sec:results} we present the results, giving the expected neutrino losses for Type I X-ray bursts and in Section~\ref{sec:discussion} we discuss the results and provide a case study on the effect that overestimating the neutrino losses has had on composition predictions. Our conclusions are given in Section \ref{sec:conclusion}. 

\section{Methods}\label{sec:methods}

\KEPLER is a one-dimensional implicit hydrodynamics code that allows for a general mixture of radiation, ions, and degenerate or relativistic electrons \citep{woosley2004}.  It provides a detailed treatment of the nuclear physics and energy generation using a large nuclear reaction network.  Reaction rates for nuclei in the mass range $A=44$--$63$ are taken from shell model calculations by \citet{fisker2001} and all other reaction rates are calculated using the Hauser-Fashbach code NON-SMOKER \citep{Rauscher2000}.  It is worth noting that there are some uncertainties inherent in the calculation of rates near the proton drip line, as discussed by \citet{fisker2001}.  Calculations include electron capture, nuclear $\beta^{+}$-decays (positron emission) and neutrino energy losses \citep{woosley2004}.

Neutrino losses during weak decays are significant, typically taking away
$\approx35\,\%$ of the energy available in a decay \citep{woosley2004}.  \KEPLER estimates the energy taken away by neutrinos during a burst using the neutrino energy loss rates of weak decay reactions from \citet{LMP} or \citet{FFN}, or using experimentally determined ground state weak strength distributions for a few light nuclei, or by taking the product of the weak decay rate and the average neutrino energy of predictions from a code from Petr Vogel \citep[see][]{woosley2004}.  \citet{woosley2004} note that since phase space heavily favours the transitions with the most energetic outgoing neutrinos, empirical strength distributions can do a fair job at estimating average neutrino energies despite the fact they cannot reliably reproduce ground state decay rates.

We computed $84$ \KEPLER models of a system that exhibits Type I X-ray bursts for initial hydrogen mass fractions of $0.2$--$0.8$, metallicity mass fractions of $0.1$, $0.02$, and $0.005$ and accretion rates of $0.3$, $0.2$, $0.1$, and $0.01$ $\dot{m}_\mathrm{edd}$, where $\dot{m}_\mathrm{edd}$ is the Eddington accretion rate, $8.8 \times 10^{-4} \frac{1.7}{(X+1)} \,$g$\,$cm$^{-2}\,$s$^{-1}$. We use the same setup as in \citet{johnston2018}: a neutron star radius of $10\,$km, a neutron star mass of $1.4\,\mathrm{M}_{\odot}$.  We start accretion on top of an iron substrate of mass $2\E{25}\,$g and use a base heating rate of $0.1\,\MeVn$.  Since the layer considered is thin compared to the neutron star radius ($\sim10\,$m) and we are only interested in the gravity of the local frame, we just use Newtonian gravity.  The local gravity corresponds to a general relativistic case of same gravitational mass but radius of $11.2\,$km.  All accretion rates, time scales, energies, and luminosities in this paper are given in the local Newtonian frame. For clarity, we have provided a table of key definitions of terms used in this paper in Table \ref{tab:defs}.

For simplicity we use a simplified composition setup that uses $^{14}$N as the only metal plus $^1$H and $^4$He.  That is, we use the $^{14}$N mass fraction synonymous for ``metallicity.''  The initial abundance of CNO isotopes is the key aspect of metals affecting the bursting behaviour.  The reader may note that in solar composition there is other metals as well, and hence in such a mix the metallicity would need to be higher to have the same CNO number abundance as in the models presented here.  Hydrogen fractions higher than $0.8$ are not observed and hard to achieve due to Big Bang Nucleosynthesis, and so are not explored in this work.  Likewise, metallicities lower than $0.005$ and higher than $0.1$ are unlikely in the systems studied here, with metallicity usually assumed to be around solar ($0.02$) so we do not explore metallicities outside those ranges.  

The accretions rates explored in this grid correspond to Cases 2--4 burning \citep[as defined by][]{keekandheger2016}, or Case III and IV burning \citep[as defined by][]{fujimoto1981}.  In these cases we expect to find hydrogen/helium rich bursts.  This is discussed in Section \ref{sec:burningregimes}.  Each run produced multiple bursts. We calculated burst energies by taking the average of the simulated bursts, excluding the first.  Uncertainties were taken as the standard deviation of the burst energies.  We exclude any model runs that produces one or fewer bursts, as these cannot give a reliable energy estimate because the first burst does not have the "chemical inertia" (ashes layer) of the later burst in a sequence \citep{woosley2004}.  We measured the neutrino flux by taking the neutrino flux of every burst from \KEPLER.  Because the neutrino fluxes of bursts in individual runs varied from burst to burst, we did not just take an average for the entire run, but instead look at the results for each burst separately.  To calculate energy generation per nucleon, the ignition column of the burst was found by assuming this was the accreted fuel since the last burst, though in our models ignition may occur above or below this point, in or above the ashes of the previous burst.  The exact point of ignition of the thermonuclear runaway is difficult to extract from a multi-zone model as it depends on the exact definition of what ignition actually is: whether one defines a temperature, convection, luminosity threshold, or similar.  Therefore we use this simple approach that is consistent with definitions used in other models which assumes that all the accreted fuel is burnt in the burst, and no residual ashes are accounted for.  We calculated the total burst energy, i.e., nuclear energy release less neutrino losses, and the total neutrino energy release for a range of \Xb values.  And \Xb was defined as the hydrogen present in the zones above the assumed ignition depth at ignition time.

\begin{table*}[h]
\centering
\caption{Definitions of key terms used in this paper \label{tab:defs}}
\begin{tabular}{llp{12cm}}
\hline
\hline
Symbol & Units & Definition \\
\hline

\Qnuc & \MeVn & \textbf{Nuclear energy generation per nucleon.} The energy released due to nuclear burning per nucleon burnt.  It is computed from the difference in mass excess per nucleon between the start and end of a burst.\\

\Xb & 1 & \textbf{Average hydrogen fraction of the ignition column.} The hydrogen fraction of the material in the ignition column as measured just before the burst ignites.  \\

$y_\mathrm{acc}$& $\mathrm{g}\,\mathrm{cm}^{-2}$ & \textbf{Accretion Column.}  Column of material that is accreted since the previous burst until the ignition of the current burst.  \\

$y$ & $\mathrm{g}\,\mathrm{cm}^{-2}$ &  \textbf{Ignition Column.} Column of material that is ignited in the burst.  For the sake of comparison with one-zone models, in this paper we define the ignition column as the same column as the accretion column.  In general, however, the ignition point may lie within the ashes from previous bursts.  See discussion in Section \ref{sec:methods}.\\

$E_\mathrm{b}$ & $\MeV$ & \textbf{Burst Energy.}  Energy released in the form of photons from the surface of the neutron star during a burst. \\

$E_\mathrm{\nu}$ & $\MeV$ & \textbf{Neutrino Energy.}  Energy released during a burst in the form of neutrinos.  Measured by summing the total neutrino energy released between the start and end of the burst. \\

$E_\mathrm{tot}$ & $\MeV$ & \textbf{Total Burst Energy.} Energy released during a burst in the form of photons and neutrinos. $E_\mathrm{tot}$ = $E_\mathrm{b}$  + $E_{\nu}$. \\

\hline

\end{tabular}
\end{table*}

\section{Results}\label{sec:results}

\subsection{Neutrino Losses in Type I X-ray Bursts}

\begin{figure}[!h]
\includegraphics[width=\columnwidth]{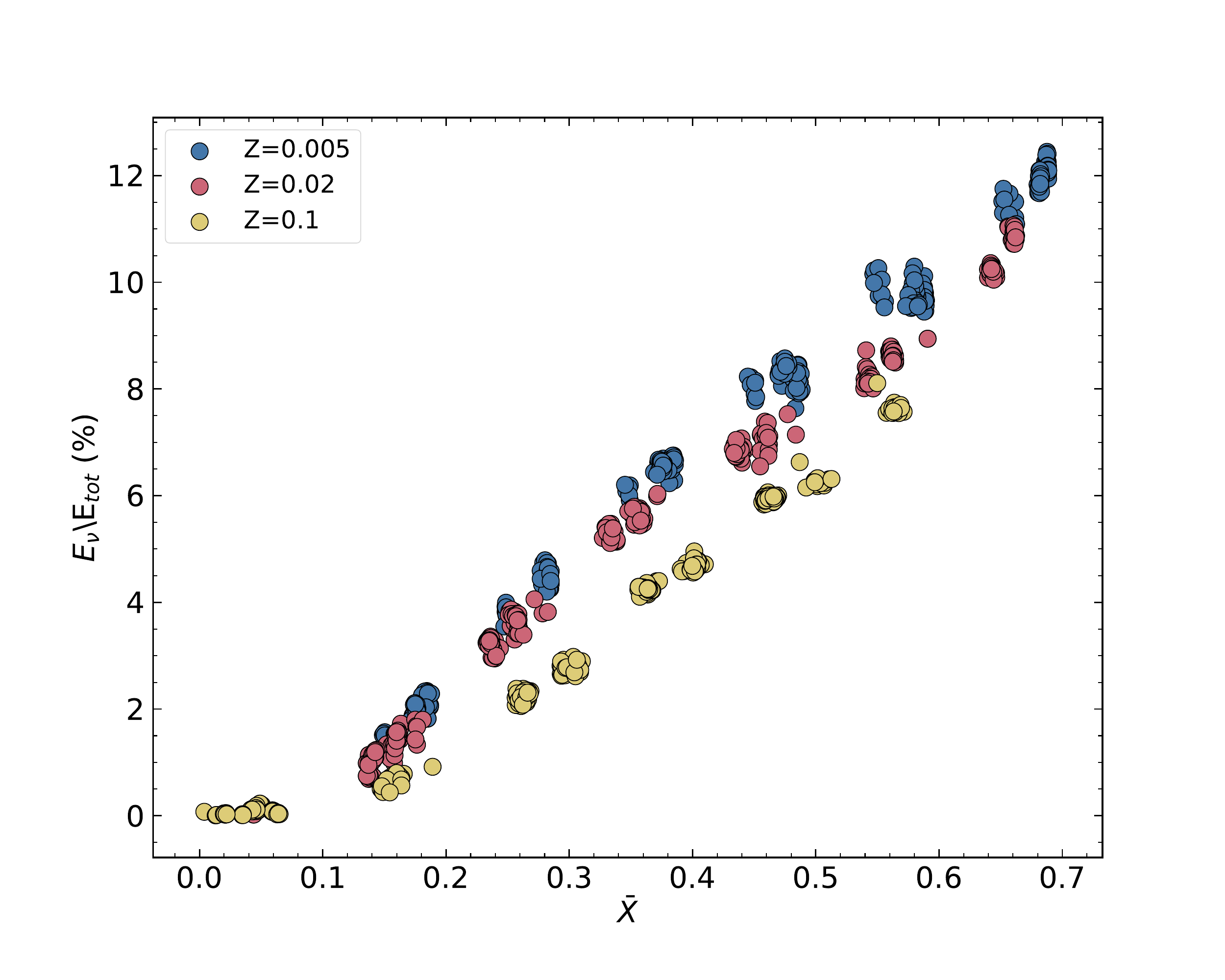}
\caption{The ratio of neutrino energy ($E_{\nu}$) to total burst energy ($E_\mathrm{tot}$ = $E_\mathrm{b} + E_\mathrm{\nu}$) for a range of initial hydrogen fractions and metallicities ($Z$).  $\Xb$ is the average hydrogen mass fraction of the ignition column.  \textsl{Yellow points} correspond to $Z=0.1$, \textsl{red points} to $Z=0.02$ and \textsl{blue points} to $Z=0.005$. \label{fig:nlosses}}
\end{figure}

We extracted the neutrino losses for each run from our multi-zone calculations and find neutrino losses range from $6.7\E{-3}\,\%$ ($\Xb \approx0$) to $14.2\,\%$ ($\Xb\approx0.75$).  See Figure~\ref{fig:nlosses}. 

When more hydrogen is present, more \textsl{rp}-process burning occurs and thus more energy is lost in the form of neutrinos from $\beta$-decays.  One can see clearly that there is never as much as $35\,\%$ energy lost as neutrinos for any amount of hydrogen present. There are differences in neutrino energy for the $3$ different metallicities explored, as we discuss in Section~\ref{sec:metallicityeffect} below.

\subsection{A New Nuclear Energy Generation Estimate}

\begin{figure*}[ht]
	\plotone{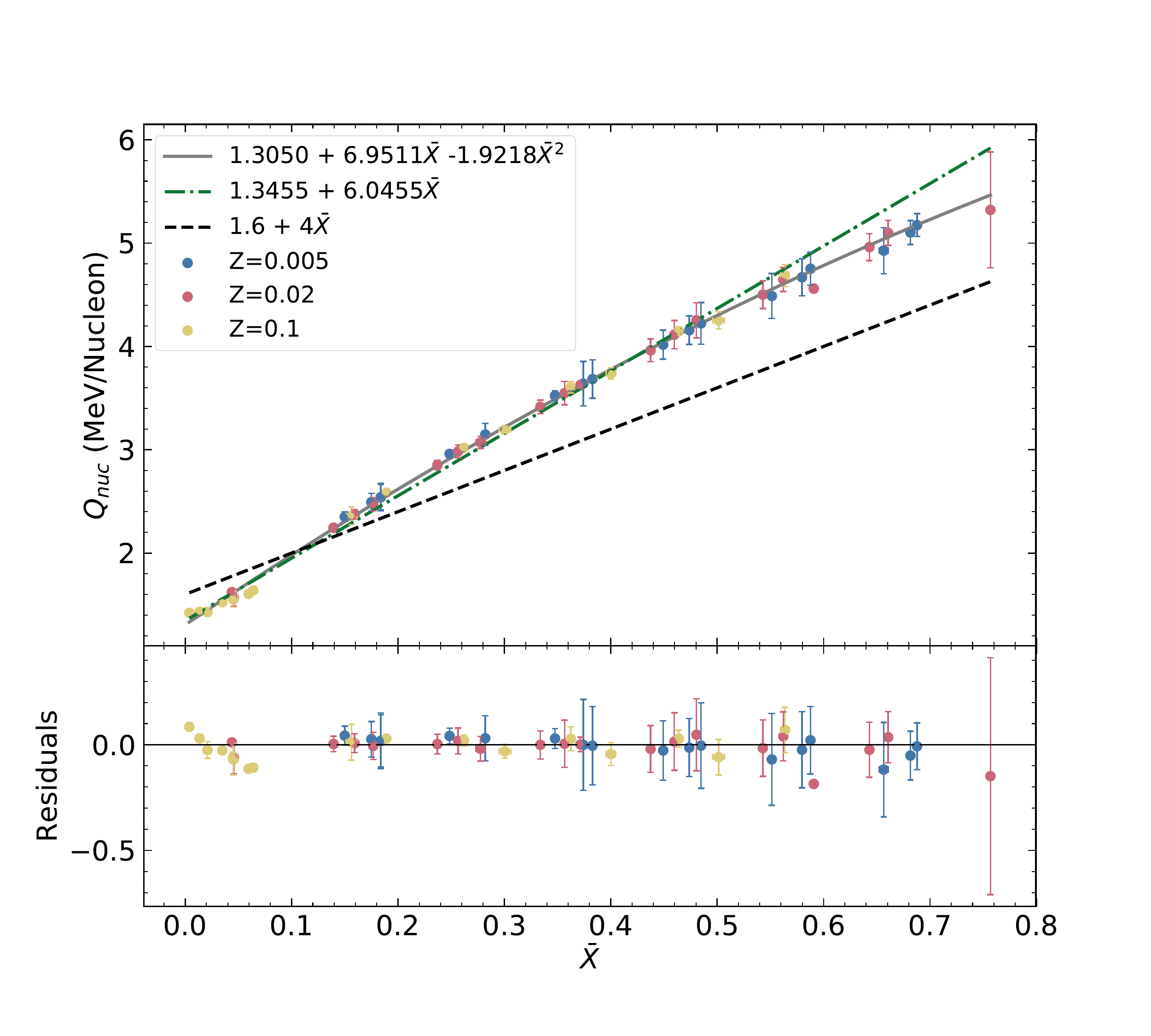}
    \caption{\KEPLER{} \Qnuc predictions (\textsl{circles}) for a range of metallicities, $Z$, and initial hydrogen fractions, $X$, compared to the approximation $\Qnuc=(1.6+4\,\Xb)\,\MeVn$ (\textsl{dashed black}).  \textsl{Yellow points} correspond to $Z=0.1$, \textsl{red points} to $Z=0.02$ and \textsl{blue points} to $Z=0.005$.  The best fit to the \KEPLER data is given by $\Qnuc=(1.31+6.95\,\Xb-1.92\,\Xb^2)\,\MeVn$ (\textsl{grey curve}).  A first order approximation is given by $\Qnuc=(1.35+6.05\,\Xb)\,\MeVn$ (\textsl{green line}).  The residuals ($Q_{\rm nuc, obs}-Q_{\rm nuc, exp}$, where $Q_{\rm nuc, exp}=(1.31+6.95\,\Xb-1.92\,\Xb^2)\,\MeVn$) are shown in the \textsl{bottom panel}.\label{fig:newrelation}}
\end{figure*}

In Figure~\ref{fig:newrelation} we show the nuclear energy generation of the bursts as a function of the average hydrogen fraction of the ignition column, \Xb.  Both the burst energy (nuclear energy less neutrinos) and the neutrino energy deviate significantly from $\Qnuc=(1.6+4\,\Xb)\,\MeVn$ due to the overestimation of the neutrino losses and reduced energy release (see Section~\ref{sec:incomplete} below). 

Fitting the multi-zone model data points using a $\chi^2$ minimisation method gives a new relation: $\Qnuc=(1.31+6.95\,\Xb-1.92\,\Xb^2)\,\MeVn$, with a reduced $\chi^2$ value of $2.32$ for $57$ degrees of freedom ($n_\mathrm{obs}-3$) and root mean square error (RMS) of $0.052\,\MeVn$, as shown in Figure~\ref{fig:newrelation}.  We require a second order polynomial to account for a trend observed in the residuals of a first order fit, $\Qnuc=(1.35+6.05\,\Xb)\,\MeVn$, which has a reduced $\chi^2$ value of $3.55$ for $58$ degrees of freedom and a RMS of $0.15\,\MeVn$.  We find there is a statistical significance in going to a second order polynomial according to the F-test \citep{rawlings2001}.  We find there is no statistical significance in going to a $3^\text{rd}$ order polynomial according to the F-test. 

The previous relation of $\Qnuc=(1.6+4\Xb)\,\MeVn$ has a reduced $\chi^2$ of $75.4$ and RMS error of $0.5\,\MeVn$, and so is a statistically poor fit to the energies predicted by our multi-zone model. Whilst the reduced $\chi^2$ values of any of the fits are not close to $1$, we note that the new relation is a marked improvement on the old relation, though still being an approximation to the true full calculation of a multi-zone model.

\subsection{Incomplete Burning of Burst Fuel}
\label{sec:incomplete}

This relation for \Qnuc corresponds to an energy release of $1.305\,\MeVn$ for fuel with no hydrogen.  This value is lower than the expected value of $1.6\,\MeVn$, which is the nuclear mass excess difference in burning $^4$He all the way to $^{56}$Fe.  On examination of some runs with low hydrogen fraction of the ignition column we found that this discrepancy is caused by two factors:
first, the reaction pathways do not burn just to $^{56}$Fe but also produce other nuclei with larger mass excess; and second, some helium burns to carbon between bursts, reducing the total yield when the burst finally ignites.

\begin{figure}[ht]
	\includegraphics[width=\columnwidth]{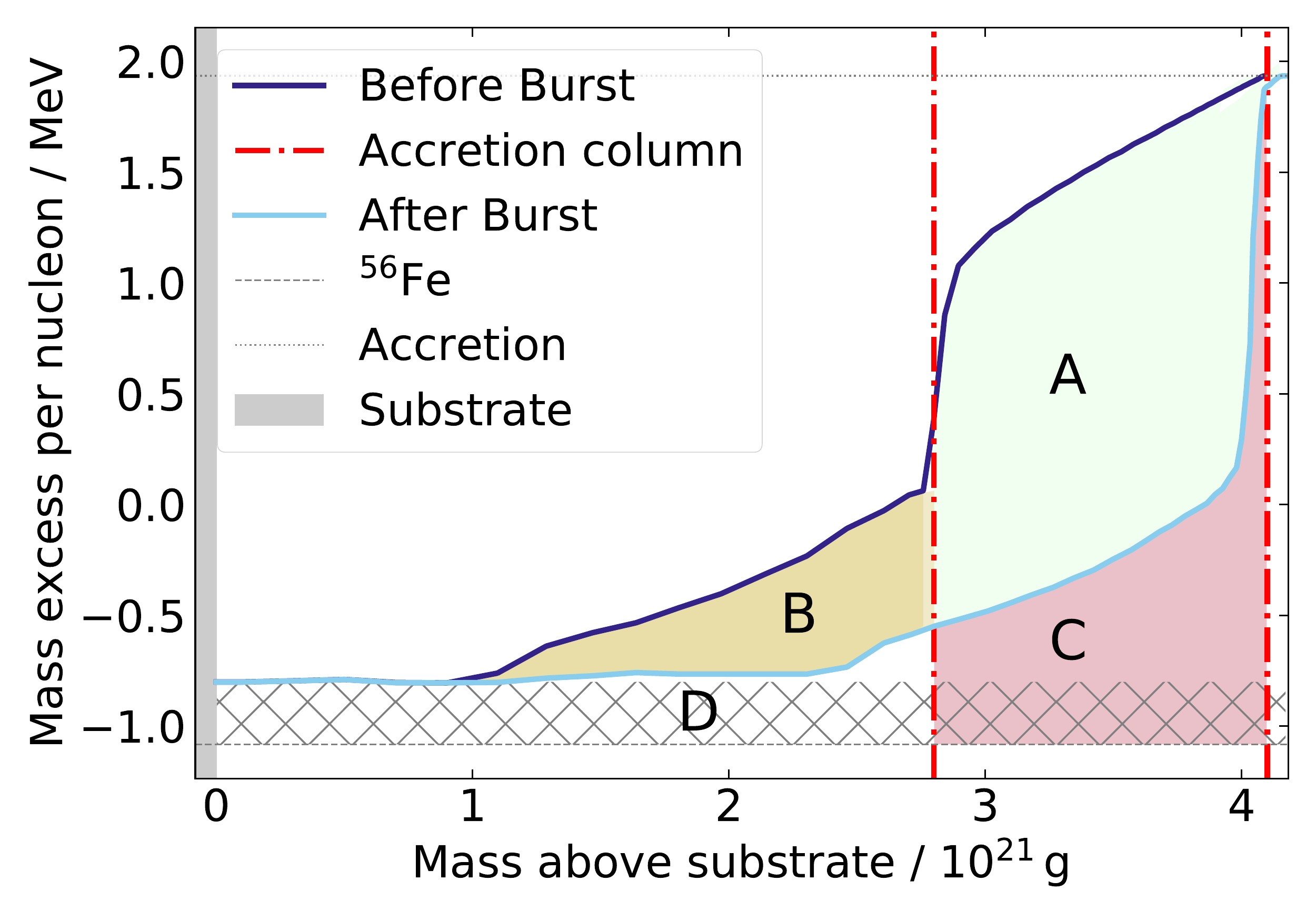}
    \caption{Mass excess per nucleon as a function of mass accreted on top of substrate just before and after a burst with very low hydrogen content in the ignition column ($\Xb=0.14$).  The dashed line corresponds to the mass excess of $^{56}$Fe, the most stable nucleus.  The labelled regions are as follows:
    \textsl{A}: Energy released in burst in the accretion column;
    \textsl{B}: Energy released in burst below the accretion column, in the ashes of the previous burst;
    \textsl{C}: Energy from fuel in the accretion column not released in this burst;
    \textsl{D}: Energy always missing in the ashes of bursts from not burning all the fuel to just $^{56}$Fe (about $0.3\,\MeVn$). \label{fig:MEplot}}
\end{figure}

Figure~\ref{fig:MEplot} shows the mass excess per nucleon just before (\textsl{dark blue curve}) and just after (\textsl{light blue curve}) a burst with low hydrogen content of the ignition column ($\Xb=0.14$).  The \textsl{dashed line} shows the mass excess per nucleon for $^{56}$Fe.  We find that the mass excess per nucleon in the ashes (\textsl{light blue curve}) is greater than that of $^{56}$Fe, the nucleus with the lowest mass excess.  Therefore the total energy produced in the burst is less than would be obtained for burning all the way to $^{56}$Fe.  Evaluating the area between the \textsl{light blue curve} and \textsl{dark blue curve}, Areas~A + B in Figure~\ref{fig:MEplot}, by integrating over the mass in the ignition column, gives us the actual total energy produced in the burst.  For the case shown, we find $\sim2.2\,\MeVn$, in agreement with  the $2.2\,\MeVn$ of our multi-zone model.  Note that for the total energy input to the model there is also a base heating of $0.1\,\MeVn$, however, that is mostly released between the bursts.

Evaluating the area between the light blue curve and the $^{56}$Fe line (Area~C), we find the energy missing from not burning all the way to $^{56}$Fe as $0.9\,\MeVn$ in the ignition column.  In contrast, if we were burning H to $^4$He and $^4$He to $^{56}$Fe we would expect to get $Q_{\rm nuc}=1.6\,\MeVn$ (helium burning to iron group) + $6.7\,\Xb\,\MeVn$ (hydrogen burning to helium) = $2.54\,\MeVn$ for the material in the ignition column.  Adding together the missing energy (Area~C) and the energy released in the burst in the ignition column (Area~A), we find $Q_{\rm nuc}\approx1.5\,\MeVn$ (burst energy) + $1.1\,\MeVn$ (missing energy) = $2.6\,\MeVn$.  This is very close to the expected value, but our numerical model will have had some extra burning, e.g., $3\alpha$ reactions during the runaway phase of the burst at the base of the ignition layer. 

\begin{figure}[ht]
	\includegraphics[width=\columnwidth]{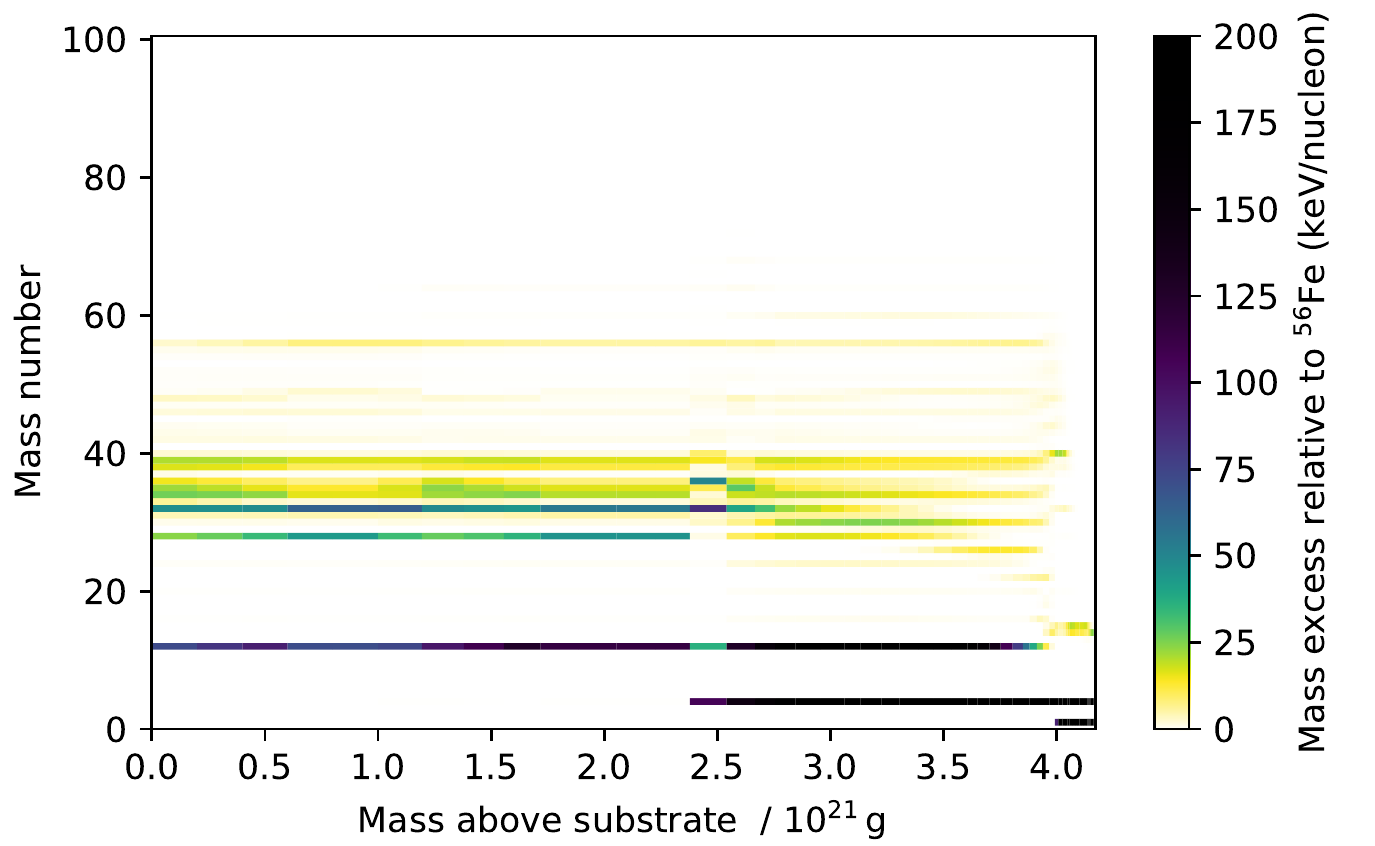}
    \caption{Mass excess per nucleon relative to $^{56}$Fe (\textsl{colour coding}) as a function of mass accreted on top of substrate (\textsl{x-axis}) and how it is distributed by mass number (\textsl{y-axis}) for a snapshot taken just after a burst with low hydrogen fraction of the ignition column ($\Xb=0.14$).  This is the same model as the \textsl{light blue curve} in Figure~\ref{fig:MEplot}.  Mass excess has been truncated at $200\,\keVn$.
    \label{fig:MEmassnumplot}}
\end{figure}

The missing energy from not burning to iron group elements is due to two factors: 1) incomplete burning of fuel leaving elements lighter than $^{56}$Fe after the burst; and 2) production of heavier elements with higher mass excess by the \textsl{rp}-process.  In Figure \ref{fig:MEmassnumplot} we show the mass excess relative to $^{56}$Fe as a function of accretion mass and how it is distributed by nuclear mass number for the ashes of the same burst as in Figure \ref{fig:MEmassnumplot}.  In this case the mass excess in the ashes is predominately in elements below $^{56}$Fe, giving strong evidence that this missing energy is caused by incomplete consumption of mass excess of the fuel, rather than excessive \textsl{rp}-process burning.

Furthermore, we found that in the cases where most of the hydrogen had been burned to helium before ignition conditions for a burst were met, there was some evidence for helium burning to carbon before the burst. The resulting  ignition column consists of up to $\approx10\,\%$ $^{12}$C, which would also reduce the total energy per nucleon available for the burst from the expected value for pure helium fuel burning to the iron group.  Burning pure carbon produces $0.6\,\MeVn$, so this would reduce the expected energy output, which could be significant for cases in which almost all the carbon burns prior to ignition \citep[e.g.,][]{keekandheger2016}.  In the specific modest case mentioned above, $0.06\,\MeVn$ were lost.

Interestingly, we find a significant fraction of the burst energy ($0.6\,\MeVn$, $40\,\%$ of the total burst energy) is released from fuel burning below the accretion column, in the ashes of the previous burst (Area~B in Figure~\ref{fig:MEplot}).  We find that an even higher fraction of fuel ($50\,\%$) does not undergo significant burning in the burst (Area~C), leaving some of this to be burnt in subsequent bursts (where it will become Area~B).  Finally, we find that in this case all bursts leave $\sim0.3\,\MeVn$ of mass excess in the ashes ($14\,\%$ of the energy released in the burst) that is not burnt in the burst, or any subsequent bursts (Area~D). In general, there is an average of 15$\%$ mass excess of the material burnt in the burst in the ashes that is leftover, and not burnt in any bursts, since the average relation is 15$\%$ less than the expected value for burning $^4$He to $^{56}$Fe. 

\subsection{Effect of Metallicity and Accretion Rate}\label{sec:metallicityeffect}

We ran models for 3 metallicites: $Z=0.005$, $0.02$, and $0.1$ and find no significant deviation from the relation in Figure~\ref{fig:newrelation}, with higher metallicities causing a larger variation in the \Qnuc value for constant \Xb. Likewise, for the 3 accretion rates explored, we find no significant deviation from the relation, with higher accretion rates causing a slightly larger scatter in the data. Whilst there is no significant deviation from the \Qnuc relation with different metallicities, we investigated the cause of the variations seen and found that bursts with lower metallicity have more \textsl{rp}-process burning, producing more neutrinos and higher \Qnuc values than bursts with higher metallicities at the same hydrogen fractions.

\begin{figure}[ht]
	\includegraphics[width=\columnwidth,viewport=18 4 648 328]{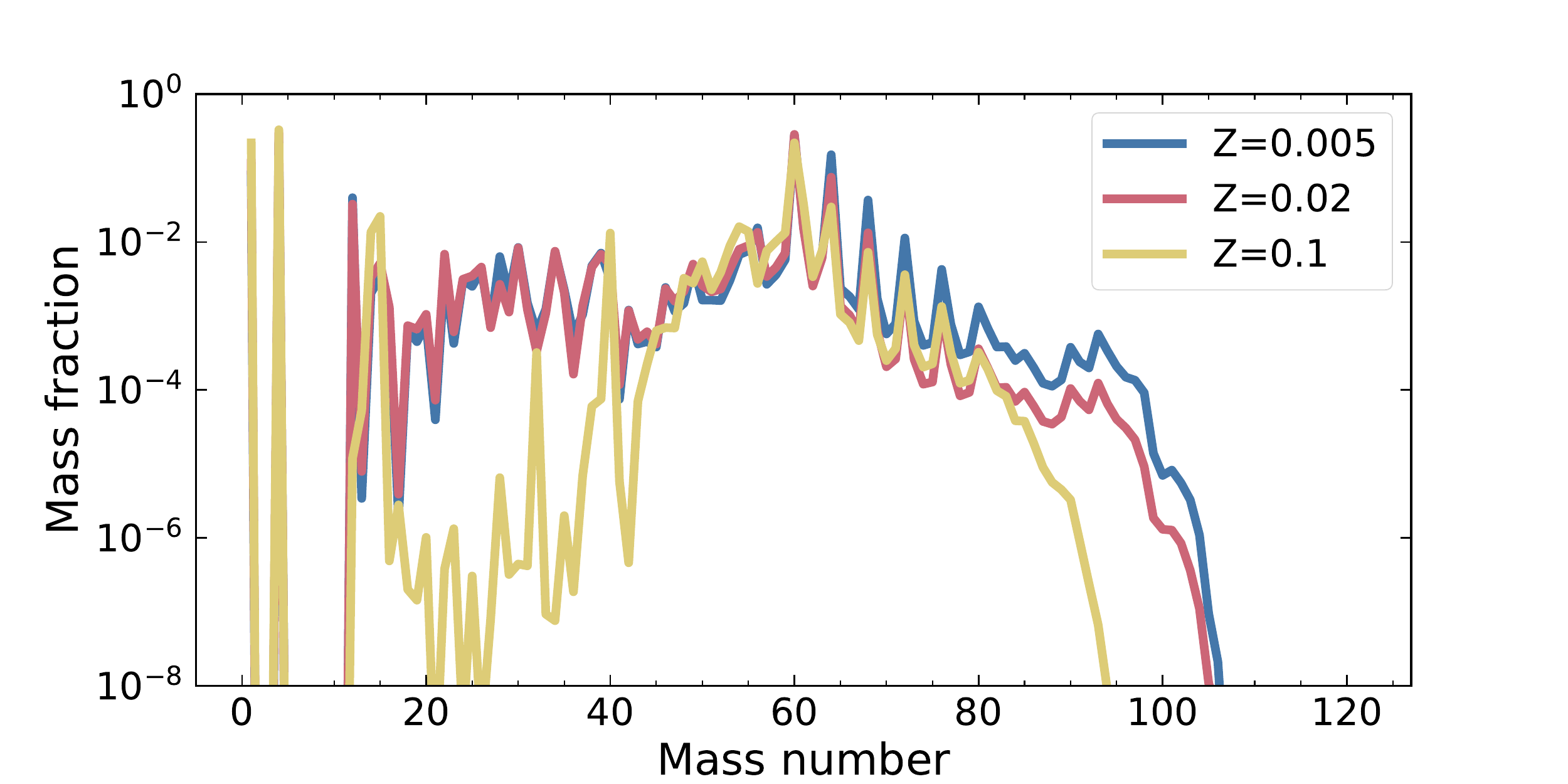}
    \caption{The mass fraction of isobars present just after a burst for $3$ models with different metallicities ($Z$) but the same hydrogen content (\Xb) of the ignition column.  The colours are the same as for Figures~\ref{fig:nlosses} and \ref{fig:newrelation}: the \textsl{yellow line} corresponds to $Z=0.1$, the \textsl{red line} to $Z=0.02$, and the \textsl{blue line} to $Z=0.005$.
    \label{fig:massfractions}}
\end{figure}

Bursts ignite with the same \Xb but different metallicities and produce different neutrino losses as a fraction of total burst energy.  We explored this further and found that, for $3$ bursts with different initial metallicities, igniting with the same \Xb but producing different neutrino losses, the distribution of heavy elements just after the burst were different, as seen in Figure~\ref{fig:massfractions}.

We find that for lower metallicity, more \textsl{rp}-process burning occurs, reaching a higher mass number, and ultimately producing more $\beta$-decays and so more neutrinos. Conversely, for higher metallicity, the \textsl{rp}-process burning stops at a lower mass number, resulting in less neutrinos produced in these cases. This result confirms that $\beta$-decays play an important role in the amount of neutrinos released during a burst, with more neutrinos released when the reaction pathways favour a greater number of $\beta$-decays, which corresponds to more prolonged \textsl{rp}-process burning.

For the purpose of this paper we note that whilst there is a trend observed with metallicity, the difference in neutrino losses is only $\approx2\,\%$ (smaller than our percentage difference in the $3$ metallicites we explored) and thus we do not include a metallicity component in any of our relations.  As discussed above, this introduces some scatter to the \Qnuc relation we have derived, but it remarkably follows a tight relation, with factors such as accretion rate and Z causing slight shifts up or down the line.

\subsection{Burning Regimes}\label{sec:burningregimes}

\citet{keekandheger2016} define $5$ distinct burning regimes based on the accretion rate and type of burning that occurs before and during a burst. In different burning regimes there are different contributions to the energetics of bursts and a single formula for \Qnuc may not hold.  In our model grid we explored three different burning regimes: $0.1$--$4\,\%\,\dot{m}_{\mathrm{Edd}}$ corresponding to He flash (stable H burning), $4$--$8\,\%\,\dot{m}_{\mathrm{Edd}}$ corresponding to stable H/He burning and $11$--$100\,\%\,\dot{m}_{\mathrm{Edd}}$ corresponding to Mixed H/He flash. (For \citet{fujimoto1981}'s burning regimes this corresponds to Case~IV pure He bursts, $\dot{m}=0.01$--$0.1\,\dot{m}_{\mathrm{Edd}}$, and Case III mixed H/He bursts, $\dot{m}=0.1$--$1.0\,\dot{m}_{\mathrm{Edd}}$).  Unfortunately, of the $24$ runs with  $\dot{m}<1\,\%\,\dot{m}_{\mathrm{Edd}}$ we studied, only $4$ were usable due to the low accretion rate and time cut off for the runs causing most of the runs to only produce one or less bursts, and thus not a reliable estimate of the energy.  As expected, the $4$ low accretion rate runs have very low \Xb of the ignition column, despite having initially high hydrogen fraction.  We have insufficient data to determine if this different burning regime causes a significant deviation from the \Qnuc relation we have developed and caution that this relation may only be valid for burning in the accretion rate range $0.1$--$1.0\,\dot{m}_{\mathrm{Edd}}$.

\section{Discussion}\label{sec:discussion}

We find that neutrino losses in Type I X-ray bursts are only significant (up to $14.2\,\%$) when there is a large amount of hydrogen present in the ignition column of the burst (mass fraction $>0.5$).  We also find that lower metallicity of the accreted fuel results in more energy released in the form of neutrinos than higher metallicity compositions.  We found that lower metallicity fuel results in more prolonged \textsl{rp}-process burning than higher metallicity fuel with the same hydrogen content, resulting in slightly more neutrinos produced.  This confirms the obvious, namely that neutrino losses are only significant at $\beta$-decays of the \textsl{rp}-process as according to \citet{wallace1981,fujimoto1987}. 

The nuclear burning processes and the resulting energy release in Type I X-ray bursts are complex, with many non-linear interaction between burning and the structure.  Therefore it is not surprising that it may not be reproduced by a simple relation that depends only on the hydrogen fraction of the ignition column.  There is no physical reason for the burst energies to be linearly proportional to the hydrogen present in the ignition column, and thus it is also unsurprising that we find a non-linear relation best fits the data.  The energy deposited in the star cannot be easily estimated by the mass excess of hydrogen and helium alone:  Depending on the reaction path different amounts of neutrinos will be carried away, and the ashes are not just pure $^{56}$Fe but do contain both heavier and light nuclei, both with large mass excess (Figure~\ref{fig:MEmassnumplot}).  Our results depend on the nuclear data being used, affecting the reaction rates and the nuclear reaction flow, and on the details and physics of the multi-zone model used (\KEPLER), affecting, e.g., mixing and transport processes.  

\subsection{Effect on Composition Predictions: A Case Study}

We quantify the overall effect of using $\Qnuc=(1.31+6.95\,\Xb-1.92\,\Xb^2)\,\MeVn$ instead of assuming $35\,\%$ neutrino losses with $\Qnuc=(1.6+4\,\Xb)\,\MeVn$ with a case study of two of the most well known accretion powered millisecond pulsars: SAX J1808.4--3658 \citep{1808discovery} and GS 1826--24 \citep{1826paper}.  \citet{galloway2006} find that the observed $\alpha$ values (where $\alpha$ is the ratio between nuclear burning energy and gravitational energy) of $\approx150$ for AX J1808.4--3658 imply that $\Qnuc\approx2\,\MeVn$.  From that they infer an average hydrogen fraction of the ignition column of $\Xb\approx0.1$.  Using $\Qnuc=(1.31+6.95\,\Xb-1.92\,\Xb^2)\,\MeVn$ ($\Qnuc=(1.35+6.05\,\Xb)\,\MeVn$) instead gives $\Xb\approx0.102$ ($0.107$), corresponding to a maximum of $10\,\%$ increase in the predicted average hydrogen fraction of the ignition column for this source. For GS 1826--24, \citet{1826paper} find the observed $\alpha$ values imply that \Qnuc $\approx 3.8 \,\MeVn$ and infer an average hydrogen fraction of the ignition column of $\Xb\approx0.55$. Using $\Qnuc=(1.31+6.95\,\Xb-1.92\,\Xb^2)\,\MeVn$ ($\Qnuc=(1.35+6.05\,\Xb)\,\MeVn$) instead gives $\Xb\approx0.403$ ($0.405$), corresponding to a $\approx 30\%$ decrease in the predicted hydrogen content of the burst fuel.

We note that the old relation fits better for low hydrogen fractions but deviates more significantly from the new formula at higher H fractions (see Figure~\ref{fig:newrelation}) so we expect to see a larger difference in fuel composition predictions for sources with higher H fraction, such as GS 1826--24.  Overall, using the correct neutrino losses in the formula for \Qnuc corresponds to a maximum energy produced of $6.34\,\MeVn$ for pure hydrogen fuel, compared to $5.6\,\MeVn$ for the old relation.  The theoretical $Q$ value of hydrogen burning by the $\beta$-limited CNO cycle, less neutrinos, is $6.2\,\MeVn$, so the new relation agrees better with classical theory.

\section{Conclusion} \label{sec:conclusion}

We have shown that neutrino losses in Type I X-ray bursts range from $6.7\E{-3}$--$14.2\,\%$ of the total burst energy, which is significantly less than the often-adopted value of $35\,\%$ during the \textsl{rp}-process.  We find that by assuming there is $\sim35\,\%$ neutrino losses in hydrogen burning during an Type I X-ray burst, the typically adopted estimate for nuclear energy generation in Type I X-ray bursts of $\Qnuc=(1.6+4\,\Xb)\,\MeVn$ significantly overestimates the neutrino losses by up to a factor of $2$.  We find that $\Qnuc=(1.31+6.95\,\Xb-1.92\,\Xb^2)\,\MeVn$, or approximately $\Qnuc=(1.35+6.05\,\Xb)\,\MeVn$, more accurately approximates the nuclear energy generation for Case~III burning \citep{fujimoto1981} as predicted by our multi-zone models.  In our models the mass excess of the fuel is incompletely extracted, reducing the energy generation for pure helium fuel from $1.6\,\MeVn$ to $1.35\,\MeVn$ as it does not burn to pure $^{56}$Fe as is classically assumed.  For the specific case discussed in detail about $0.3\,\MeVn$ ($14\,\%$ of the initial mass excess relative to iron) remains in the ashes.  Interestingly, we found evidence for a significant fraction of energy ($\approx40\,\%$) released in a burst coming from burning below the accretion depth, in the ashes of the previous burst.  We also found that the amount of energy carried away by neutrinos does noticeably depend on metallicity even for the same hydrogen fraction of the ignition column.  This  indicates that for lower metallicity there is stronger \textsl{rp}-process burning and so more neutrinos released by a larger number of $\beta$-decays or reaction pathways with higher-energetic neutrino emission, than for cases with the higher metallicity. 

\acknowledgments

We thank Andrew Cumming and Ed Brown for helpful discussions.  AG acknowledges support by an Australian Government Research Training (RTP) Scholarship.  AH has been supported, in part, by the Australian Research Council through a Future Fellowship (FT120100363) and by TDLI though a grant from Science and Technology Commission of Shanghai Municipality (Grants No.16DZ2260200) and National Natural Science Foundation of China (Grants No.11655002).

\software{\KEPLER \citep{woosley2004}, NON-SMOKER \citep{Rauscher2000}}

\bibliographystyle{apj}
\bibliography{bibfile}




\end{document}